\documentclass[conference]{IEEEtran}
\IEEEoverridecommandlockouts

\usepackage{cite}
\usepackage{amsmath,amssymb,amsfonts}
\usepackage{algorithmic}
\usepackage{graphicx}

\usepackage{textcomp}
\usepackage{xcolor}
\usepackage{tikz}
\usepackage{amsmath}
\usepackage{filecontents}
\usepackage{graphicx}
\usepackage[ruled,vlined]{algorithm2e}
\usepackage{import}
\usepackage{amsthm}
\usepackage{subcaption}
\usepackage{enumitem}
\usepackage{wrapfig}
\usepackage{stfloats}
\def\BibTeX{{\rm B\kern-.05em{\sc i\kern-.025em b}\kern-.08em
    T\kern-.1667em\lower.7ex\hbox{E}\kern-.125emX}}
\begin{document}

\author{
    \IEEEauthorblockN{Tasnia Ashrafi Heya\IEEEauthorrefmark{1}, Abdul Serwadda\IEEEauthorrefmark{1},
    Isaac Griswold-Steiner\IEEEauthorrefmark{1}, Richard Matovu\IEEEauthorrefmark{2}}
    \IEEEauthorblockA{\IEEEauthorrefmark{1}Texas Tech University, Lubbock, Texas, USA\\
    \{tasnia.heya, abdul.serwadda, isaac.griswold-steiner\}@ttu.edu}
    \IEEEauthorblockA{\IEEEauthorrefmark{2}Gannon University, Erie, Pennsylvania, USA\\
    \ matovu001@gannon.edu}
}

\newcommand*{\sectiondir}{Sections/}

\title{Using wrist movements for cyber attacks on examination proctoring
}

\maketitle

\begin{abstract}
Multiple choice questions are at the heart of many standardized tests and examinations at academic institutions allover the world. In this paper, we argue that recent advancements in sensing and human-computer interaction expose these types of questions to highly effective attacks that today's proctor's are simply not equiped to detect. We design one such attack based on a protocol of carefully orchestrated wrist  movements combined with haptic and visual feedback mechanisms designed for stealthiness. The attack is done through collaboration between a knowledgeable student (i.e., a mercenary) and a weak student (i.e., the beneficiary) who depends on the mercenary for solutions. Through a combination of experiments and theoretical modeling, we show the attack to be highly effective. The paper makes the case for an outright ban on all tech gadgets inside examination rooms, irrespective of whether their usage appears benign to the plain eye.
\end{abstract}

\begin{IEEEkeywords}
Wearables, Gesture-based interaction
\end{IEEEkeywords}

\IEEEpeerreviewmaketitle

\section{Introduction}
In this paper, we show that human-machine interactions made possible by a wrist-worn, sensor-enabled device could be leveraged to stealthily cheat an examination. We particularly focus on a smart watch, showing how a methodical combination of its motion sensing, haptic feedback and visual feedback mechanisms can be used by students to cheat an exam. We alternately use the term “attack”, to refer to the mechanism of cheating the exam. Specifically, we use the example of  multiple choice questions to showcase an attack centered on a simple protocol through which answers are encoded using carefully selected gestures or pen movements. 

These pen movements are executed by a person having knowledge of the exam's subject matter (e.g., a student who is well prepared for the exam) and then communicated to the receiving entity (i.e., the unprepared student) using a visual or haptic channel that is carefully tuned to evade detection. In the rest of the paper, we will use the terms {\it{mercenary}} and {\it{beneficiary}} to respectively refer to these two students.We show the attack to be highly effective and stealthy. The contributions of the paper are summarized below: 

{\bf{(1) Design of an attack that leverages sensor-enabled wearable devices to enable cheating during examinations:}} We present a new paradigm of attack that could be used by students to stealthily undertake cheating during examinations comprised of multiple choice questions. The attack exploits the myriad sensing and feedback functionalities available in today's mobile and wearable devices as a vehicle to both encode and communicate solutions to the exam. 

{\bf{(2) Evaluation of the dynamics of the attack:}} Through a series of human subjects experiments, we rigorously evaluate the performance of the attack and tune aspects of it to maximize its impact. As a channel to communicate answers back to the beneficiary, we formulate and evaluate visual and haptic based schemes that are designed to evade detection by the proctor and (or) nosy students.   

{\bf{(3) Modeling the attack:}} 
To characterize the attack beyond the scope of our human subjects experiments, we develop a theoretical model that more generally expresses the impact of the attack. The model captures the connection between the probability of passing an exam to key variables such as the accuracy of gesture recognition, the academic capabilities of the mercenary recruited by the beneficiary and the number of answer options provided per question among other variables. 
 \section{Related Research}
\label{related-1}
The paper with the greatest similarity to our own work is perhaps that by Migicovsky et al. \cite{migicovsky2014outsmarting}. Migicovsky et al. developed a system that demonstrates how students could collaboratively cheat by using smartwatches to vote on the answer they believe to be correct and request the most common answer for a particular problem via their watch. 

{\it{There are at least two fundamental differences between the work by Migicovsky et al. and our own work.}} First, Migicovsky {\it{et al.}} did not perform any experiments to evaluate their proposed attack. The authors only described their proposed attack and developed an app to support the discussion on the features of the attack. Without any form of experimental or theoretical evaluation on their attack, it is not clear how much of a threat the attack would be in practice. 

An even more significant difference between Migicovsky {\it{et al.}}'s work and our work lies in the mechanisms of the attacks. The attack hypothesized by Migicovsky et al. requires users to type on the watch as they choose questions and specify answers that are then shared with collaborators. Proctors in this day and age would be very unlikely to allow a student to get onto their watch and type content or tap buttons intermittently during the execution of an exam.   
By virtue of leveraging wrist motions as the attack vehicle, we technically study a significantly different threat vector from that by Migicovsky {\it{et al.}}

A more recent research paper by \cite{wong2017assessing} evaluated the usability of smartwatches by students aiming to cheat during an exam. The authors developed a smartwatch app that allowed students to view text and images on the device while taking an exam. Using quantitative and qualitative methods the authors solicited opinions from students regarding the use of a smartwatch for cheating.  
The fact that the work in \cite{wong2017assessing} is focused on interviewing students about the general use of watches for cheating puts it apart from our research given that we design and evaluate a concrete attack.

\begin{figure*}[h]
\centerline{
\includegraphics[width=0.82\textwidth,height=4.13cm]{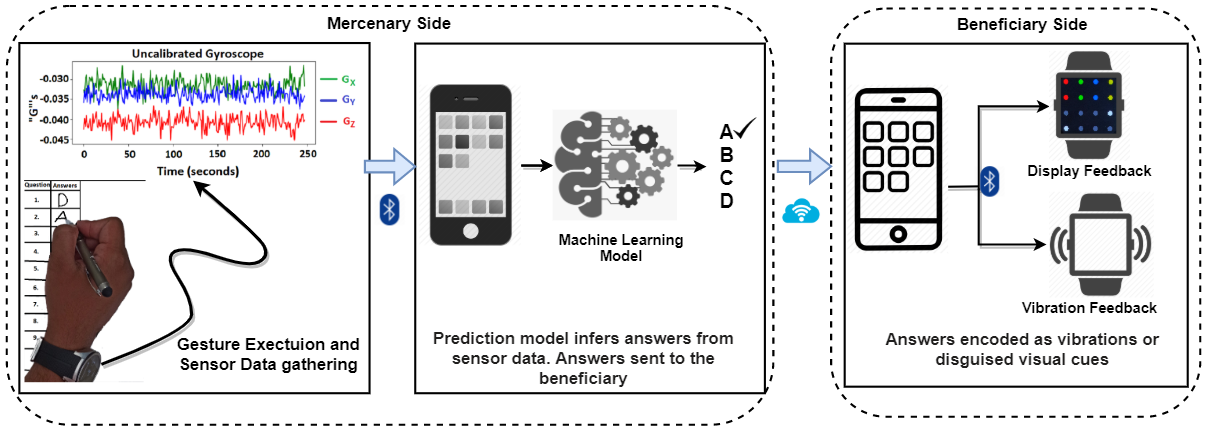}}
\caption{Overview of attack process. 
}
\vspace{-0.35 cm}
\label{fig:threatmodel}
\end{figure*}

\section{Design, Implementation and Evaluation of the Mercenary Component}
\label{attack-2}

\subsection{Attack design}
\label{attack-1}
The attack is centered on wrist motions that occur during the writing of {\it{methodically selected}} English characters. Each character maps to an answer. A sequence of characters are carefully interleaved with pauses during the attack. For illustration, we assume the case of an exam for which the answer options to each question are ``A'', ``B'', ``C'' and ``D''. Take the case of a question on this exam whose answer is ``A''. To implement our attack, the mercenary first reads the question and determines that the answer to this question is ``A''. The mercenary then has the option to either immediately trigger the attack and send this answer to the beneficiary, or to continue on to find the answer to the next question so as to accumulate a sequence of answers. Assuming the mercenary decides to send the answer to the first question, he/she executes the attack using the following three steps:

\begin{enumerate}[label=\roman*.]
 \item initiate the attack by holding the hand still on top of the table for $(t_1\pm \varepsilon)$ seconds (we also call this an {\it{opening}} pause). 
 \item write the letter ``A'' on the paper (or mimic the writing of letter ``A'')\footnote{The mercenary is free to write on the exam sheet or on scratch paper, or to simply pretend to write by tracing out a character.}, and,
 \item hold the hand still on the table to execute another pause of $(t_2\pm \varepsilon)$ seconds (we also call this a {\it{closing}} pause).
\end{enumerate}
For this research we have set $t_1=12$ seconds, $t_2=5$ seconds and $\varepsilon=2$ seconds following trial experiments in which users tried out different options for these parameters. 



After executing the closing pause, the mercenary is no longer required to stick to the protocol, and may move the hands freely, flip the pages of the exam, undertake calculations, etc. The system ignores all these actions until the mercenary initiates another opening pause that signals the start of the second answer. On completing this pause, the mercenary writes the character representing the second answer, and then executes the closing pause similarly to what was done with letter ``A''. This process continues until the end of the exam, with the questions answered in sequential order. Because data could be delayed or reordered during transit across the network, answers sent out to the beneficiary ought to have meta information (e.g., time-stamps or question numbers) to enable realigning the order of answers if needed. 

\subsection{Data collection experiments for evaluation of mercenary component}
\label{data-1}

Data collection for the mercenary module was done using a smart watch app on the Android platform. The app recorded gyroscope and accelerometer\footnote{Experiments showed accelerometer data to not perform well for our attack, so we do not reference it any further and instead focus on gyroscope data going forward.} data that was then sent via bluetooth to a smart phone paired with the watch. All watches used for our experiments were either LG Urbane (W200) or Fossil Gen 5 (FTW 4025). The watches were set to sample sensor data at a rate of 60 Hz. To execute the attacks, we recruited 15 mercenaries each of whom wore a smart watch paired with a phone. Each mercenary was given a detailed description of the attack protocol and allowed to practice before beginning attack execution. Practice times were on average about 20-30 minutes before mercenaries were ready to participate in the experiment. 

The data collection experiment varied depending on whether training or testing data was being collected. To collect training data, mercenaries simply carefully wrote each of the characters supported by our attack. 
For example, to collect training data for a character such as ``A'', the mercenary wrote the letter ``A'' immediately after the press of a start button on the interface. On completion of writing the letter, a stop button was immediately pressed so gyroscope data that precisely delimits this letter was saved. This process was repeated 29 more times for a total of thirty samples of letter ``A''. Other characters were treated similarly (details of the full range of characters that we evaluated will be discussed in Section \ref{perf-1}). Collecting training data in this way enabled us to cleanly capture the data for each character without the pause portions of the protocol complicating character segmentation and potentially adding noise that could adulterate the trained model. 

To collect testing data, each mercenary followed the full attack process (earlier described in Section \ref{attack-1}), writing answers, and pausing per specifications. Mercenaries wore the watch on the hand being used to write, meaning that this watch both recorded the wrist gestures and provided time measurements needed for the protocol. In total, each of the 15 mercenaries wrote 50 answers during the attack (an equivalent of solving 50 questions on an exam).  


\subsection{Classification framework} 
\label{classn-1} The framework to classify characters (or exam answers) begins with the detection of opening and closing pauses, followed by the classification of sensor data located between these two pauses into one of the answer classes. Pause detection uses a threshold-based approach. In particular an instantaneous gyroscope measurement $g_x$ is flagged as potentially being part of pause if $-T_h\leq g_x\leq T_h$, where $T_h$ is a mercenary-specific threshold. $T_h$ obtained from data collected from a training dataset of pauses provided by the mercenary. When a series of consecutive points meet this condition, they are compared to the criteria for the closing and opening pauses (recall Section \ref{attack-1}) and marked as such if they meet this criteria. Otherwise they are ignored.  

\begin{table}[]

    \caption{List of temporal and spectral features used to train our classifiers}
\begin{center}
\resizebox{\columnwidth}{!}{

    \begin{tabular}{|c|c|}
    \hline
    \textbf{Type}                                                & \textbf{Feature Names}                                                                                                                                                                                                                                                                                                                                                                                                          \\ \hline
    \begin{tabular}[c]{@{}c@{}}Temporal \\ features\end{tabular} & \begin{tabular}[c]{@{}c@{}}Mean, Standard Deviation, Interquartile Range, Absolute Energy,\\ Mean Absolute Deviation, Standard Error of the Mean,\\ Mean change, Autocovariance, Longest Strike Above Mean,\\ Variance, Absolute Sum of Changes, Kurtosis, Sample Entropy,\\ Autocorrelation, Mean absolute change, Sum, Skewness, Quantile,\\ Median, Longest Strike Below Mean and Complexity invariant distance\end{tabular} \\ \hline
    \begin{tabular}[c]{@{}c@{}}Spectral\\ features\end{tabular}  & \begin{tabular}[c]{@{}c@{}}Spectral Centroid, Spectral Flatness, Spectral Kurtosis,\\ Spectral Skewness, Spectral Decrease,Spectral Spread,\\ Spectral Rolloff, Spectral Slope\end{tabular}    \\ \hline                                                                                                                                                                                                                                 
    \end{tabular}}

    \label{features-1}
\end{center}
\vspace{-0.55 cm}
    \end{table}

Following pause detection, gyroscope data located between an opening and closing pause is input to the machine learning module which infers which character the data maps to.  For each of the $X$, $Y$ and $Z$ components of gyroscope data, the module extracts 32 features. These features are a mix of frequency domain and time domain features (see Table \ref{features-1} for the list of features). Thus a total of 96 features are extracted from each character (i.e., from the full window of data between the {\it{opening}} and {\it{closing}} pauses). We classified this data with two classifiers: a random forest (with 60 trees) and logistic regression. For classification, we used the default parameters in python's Scikit learn library for all classifiers as we found them to perform well. 

\subsection{Performance of classification framework}
\label{perf-1}
The pause detection module detected and classified all opening and closing pauses correctly for all mercenaries (i.e., 100\% accuracy). In the bulk of this section, we describe the character classification performance which had a much more complex dynamics.

\subsubsection{Exploratory evaluation of attack symbols}To prepare the ground for presentation of the classification performance of the character classification portion of the framework, we first make the distinction between what we refer to as {\it{answer options}} and {\it{attack symbols}} that represent the answer options. Typically the four {\it{answer options}} in a multiple choice exam are $A$, $B$, $C$ and $D$ for the first, second, third and forth answer choices respectively. The {\it{attack symbols}}, which {\it{might}} or {\it{might not}} be the same as the  {\it{answer options}}, are the characters which the mercenary writes (or pretends to write with simulated pen strokes) to represent the {\it{answer options}}. The data consumed by the machine learning algorithms is in essence the sensor data generated by the mercenary's wrist motions when writing these  {\it{attack symbols}}.

\begin{table}[ht]
\caption{Results from clustering of characters using the K-means algorithm. The results show that B, C, E, and A are in 4 different clusters most of the time, supporting their usage for our attack.}
\begin{center}
\resizebox{\columnwidth}{!}{
\begin{tabular}{|c|c|c|c|c|}
\hline
\textbf{User} & \multicolumn{4}{|c|}{\textbf{Data Clusters}} \\
\cline{2-5} 
\textbf{} &  \textbf{\textit{Cluster 1}}& \textbf{\textit{Cluster 2}}& \textbf{\textit{Cluster 3}} & \textbf{\textit{Cluster 4}}\\
\hline
1 & {\bf{B}},H,J,O,q,S,y,Z,8 & {\bf{C}},K,m & {\bf{E}},I,W & {\bf{A}},X \\
\hline
2 & {\bf{B}},I,J,m,O,q,S,8 & {\bf{C}},X & A,D,{\bf{E}},H,W,y,Z & K \\
\hline
3 & {\bf{B}},K,X,Z & {\bf{C}},J,m,O,q,S,W,y,8 & {\bf{E}},H,I & {\bf{A}} \\
\hline
4 & {\bf{B}},I,J,K,W,X & {\bf{C}},q & {\bf{E}},O,S,Z,8 & {\bf{A}},H,m,y  \\
\hline
\end{tabular}}
\label{cluster-1}
\end{center}
\vspace{-0.25 cm}
\end{table}

In selecting which {\it{attack symbols}} to be used by our attack, we first conducted an exploratory investigation of the classification performance of the {\it{answer options}} $A$, $B$, $C$ and $D$ as {\it{attack symbols}} given their usability benefits. For this investigation, we collected a separate exploratory dataset early on in the study  from a subset of mercenaries. We found $B$ and $D$ to be very frequently confused with each other (results not shown due to space limitations), likely due to their similar shape. We decided to drop ``D'' as an attack symbol and keep the others.

\subsubsection{Definitive set of attack symbols}To find a replacement for ``D'' which wont get confused with any of the 3 confirmed attack symbols, we went back to the drawing board and analyzed the alphabets A-Z and the numbers 0 to 9. This analysis involved two steps: a visual selection in which we discarded a character if it had another character having a visually similar shape to it, and, a clustering process in which we split the remaining characters into 4 clusters using the K-means algorithm. In the visual step, we dropped a wide range of characters and kept only 19. To give some insight into our selection criteria, the following are examples of characters which we dropped in favor of others at this visual stage. We chose ``E'' and dropped ``F'', chose ``Z'' and dropped ``2'', chose ``S'' and dropped ``5'', chose ``I'' and dropped ``1'', etc. 

Having dropped a subset of the characters, we had each of our mercenaries write the remaining symbols while we captured the associated gyroscope data. We then extracted the earlier described 96 features and clustered the data using K-means with K=4. Table \ref{cluster-1} shows the results from this clustering stage for 4 representative mercenaries in our exploratory dataset. Observe that ``B'' and ``C'' are always in clusters \#1 and \#2 respectively while ``A'' is in cluster \#4 75\% of the time. We observed similar patterns for other sets of mercenaries (results not shown here). The fact that these 3 characters are always in separate clusters, supports their choice as attack symbols and validates the separability. For the fourth symbol, we naturally had to focus on cluster \#3 since it does not contain any of our 3 chosen characters. Observe that ``E'' is consistently located in this cluster, making it our final choice of attack symbol. The final set of attack symbols used in our fully fledged data collection experiment (described in Section \ref{data-1}) were A, B, C and E to respectively represent the answer options A, B, C and D. This is a highly usable set that should only provide a minor headache to the mercenary (only letter E would be unusual). 

\begin{figure}[ht]
\centerline{
\includegraphics[width=0.45\textwidth]{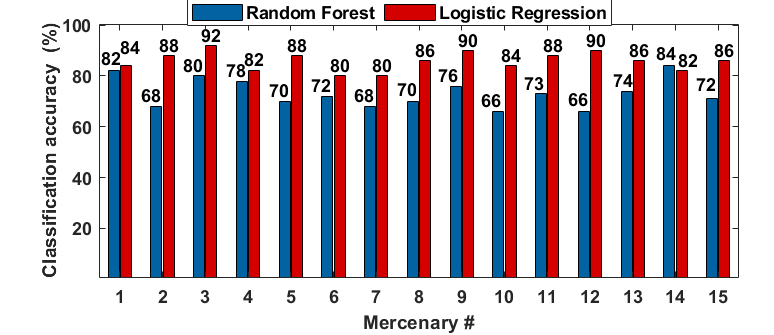}}
\caption{Classification performance of all 15 mercenaries.}
\label{fig-1}
\end{figure}

\subsubsection{Attack results for definitive set of attack symbols}
\begin{wrapfigure}{l}{0.32\textwidth}
\centerline{
\includegraphics[width=0.32\textwidth]{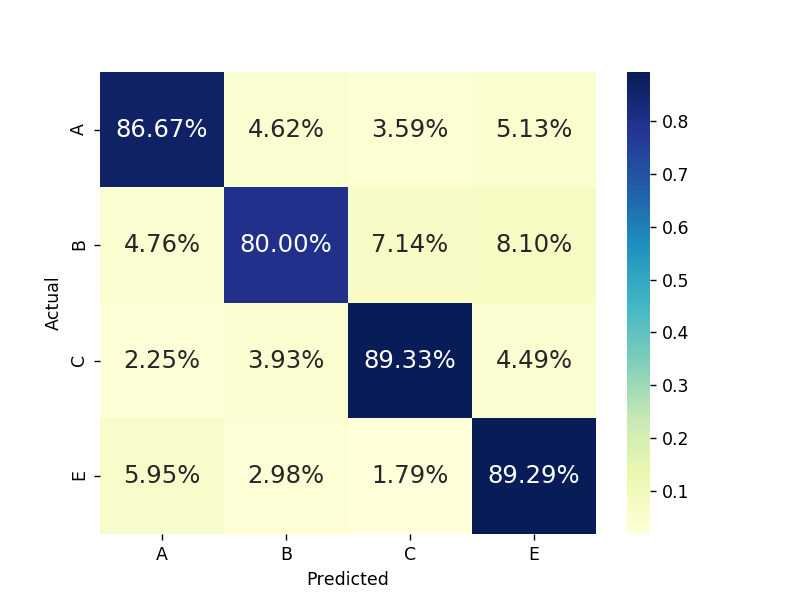}}
\caption{Confusion matrix for classification performance of 15 mercenaries.}
\label{fig-2}
\end{wrapfigure}

Figures \ref{fig-1} and \ref{fig-2} summarize the performance of the attack when we used this final character-set. Figure \ref{fig-1} shows that the Logistic Regression classifier always performed best, achieving 80\% or higher accuracy for all 15 mercenaries. The figure reveals that the Random Forest performed worse, but still had an accuracy of 70\% or more for 11 of the 15 mercenaries. 

To gain insights into the distribution of mis-classifications we show the confusion matrix obtained with the Logistic regression classifier in Figure \ref{fig-2}. We only show the confusion matrix for the best performing classifier due to space limitations. The figure is an average of the confusion matrices obtained with each of the 15 mercenaries. The figure confirms that the character ``E'' provided a boost, as no single character is causing a disproportionate amount of mis-classifications. The average accuracy rose to 86.3\% with this new character-set. A similar trend was seen with the Random Forest classifier (results not shown here). 

Overall, an average accuracy of close to 90\% for the best classifier suggests a highly effective attack. That said, this answer-level (as opposed to exam-level) behavior of the attack only paints a limited picture of its behavior. In Section \ref{model-1} (Page \pageref{model-1}), we more rigorously characterize the attack as we provide more context as to how  a classification performance in the above cited range relates to the beneficiary's odds of passing the exam.

\section{Design, Implementation and Evaluation of the Beneficiary Component}
\label{benef-1}

We designed and studied two approaches for answer retrieval by the beneficiary: a haptic approach and a visual approach. Below, we describe how we designed these approaches as well as how they performed in our experiments.  

\subsection{Haptic answer retrieval mechanism} 
\begin{wrapfigure}{l}{0.32\textwidth}
    \centering
    \includegraphics[width=0.31\textwidth]{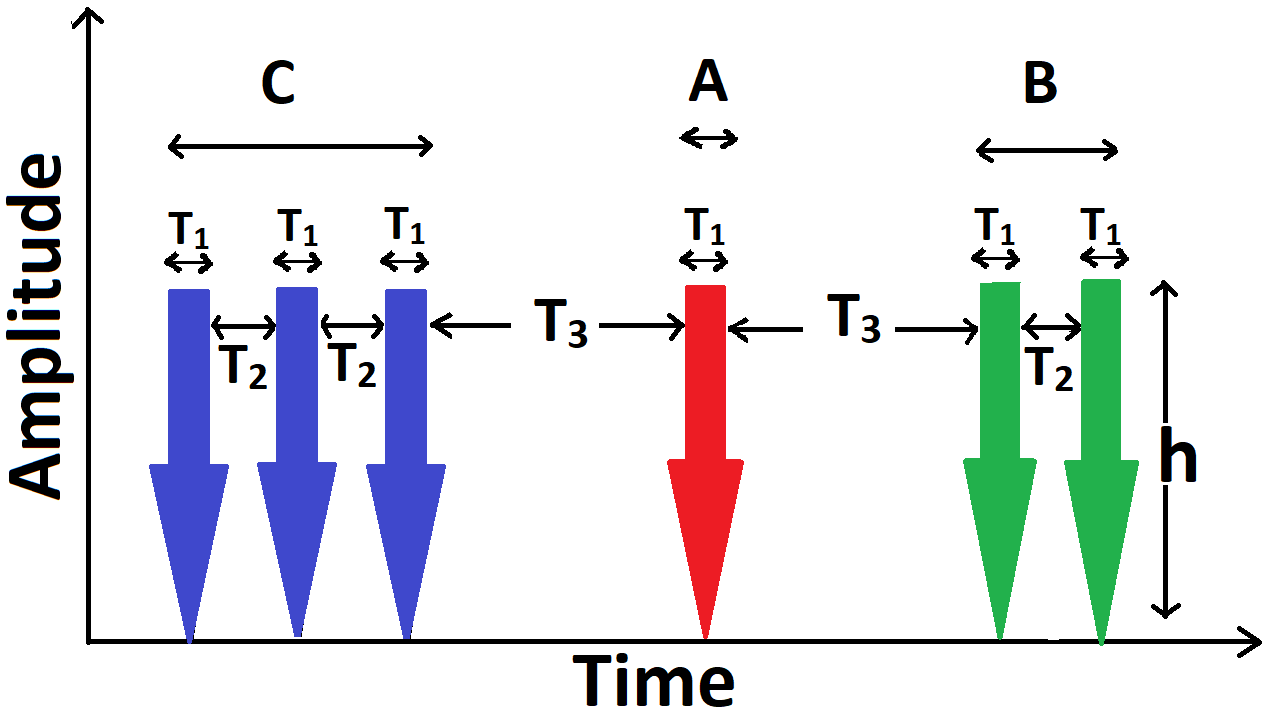}
    \centering
    \caption{Vibration-based system for answer feedback to the beneficiary. Each arrow represents a single vibration. The example represented on the figure is for the answer sequence C, A, B.}
    \label{vib-1}
    \vspace{-0.25 cm}
\end{wrapfigure}
\subsubsection{Design and implementation} The haptic approach encodes the answers as vibrations. The {\it{answer options}} A, B, C and D, are respectively represented by one, two, three and four vibrations within a cluster. Figure \ref{vib-1} illustrates this approach for an answer sequence of 3 answers, C, A and B. The first three arrows represent three vibrations and hence the answer C. The definitions of the other two (A and B) follow the same idea. Across an exam, the cluster numbers represent the question being answered (i.e., the $i^{th}$ cluster represents the $i^{th}$ question on the exam). The approach relies on 4 parameters: $T_1$ (duration of an individual vibration), $T_2$ (time interval between vibrations in a cluster), $T_3$ (time interval between clusters) and $h$ (amplitude/strength of the vibrations). We discuss our settings for these parameters in the following subsection. 

\subsubsection{Performance evaluation} 
A key aspect of our performance evaluation was how to configure the 4 parameters without compromising the attack. The following considerations guided our decisions: (1) $T_3$ ought to be markedly distinct from $T_2$ if the end-user is to discern a new cluster from a vibration within a cluster, (2) $T_1$ has to be long enough for the user not to miss a vibration when it happens, (3) $T_2$ should not be too short that the user fails to distinguish between two consecutive vibrations within a cluster, (4) $h$ has to be of such magnitude that the user is guaranteed to feel the vibrations, but it should not be too strong that the proctor hears the audio from the vibrations, (5) $T_3$ is not within the full control of the software developer. For example, if the mercenary takes a very long time to answer a question, $T_3$ could get arbitrarily long. One can only fix its minimum value to avoid confusion with $T_2$. 

\begin{wrapfigure}{l}{0.30\textwidth}
\centering
\includegraphics[width=0.30\textwidth,height=0.24\textwidth]{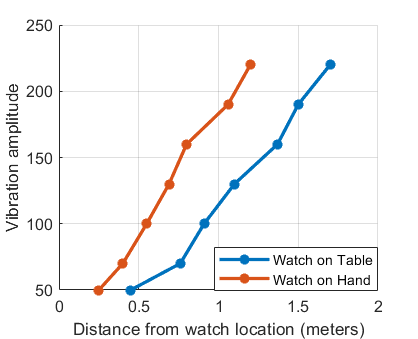}
\caption{Distance at which smart watch vibrations of various amplitudes get detected by a third party in the room.}
\label{vib-2}
\end{wrapfigure}

Due to space limitations we only provide details of our sensitivity analysis on the dynamics of $h$ since audio generated based on this parameter has a very strong connection with students potentially being caught by the proctor. 

For the other parameters, we only point out here that we ultimately set $T_1=200 ms$ and $T_2=1 second$ as these values enabled all our 15 participants to reliably delineate between individual vibrations. We set the minimum value of $T_3$ as 45 seconds. With these settings all our participants successfully retrieved all the answers. 
Figure \ref{vib-2} shows results of our sensitivity analysis on $h$. 

The figure reveals that the distance at which the vibrations are heard, grows about linearly as the amplitude increases towards 250 units. Its also clear from the plot that wearing the watch on the wrist will always be the safer strategy. In our experiments we found an amplitude of 70 units to work well as participants detected all vibrations and got answers correct under this setting with the other variables set to the earlier described values. Assuming the live attack uses this value of 70 units, the plot reveals that a beneficiary wearing the watch on the wrist would need to sit further than 0.5 meters from the proctor's location. If the classroom arrangement is so packed and such an amount of room cannot be found, the beneficiary might then have to opt for the visual feedback system (discussed next).  
 
\subsection{Visual answer retrieval mechanism} 
\subsubsection{Design and implementation}  

\begin{wrapfigure}{l}{0.23\textwidth}
    \centering
    \includegraphics[width=0.24\textwidth,height=0.23\textwidth]{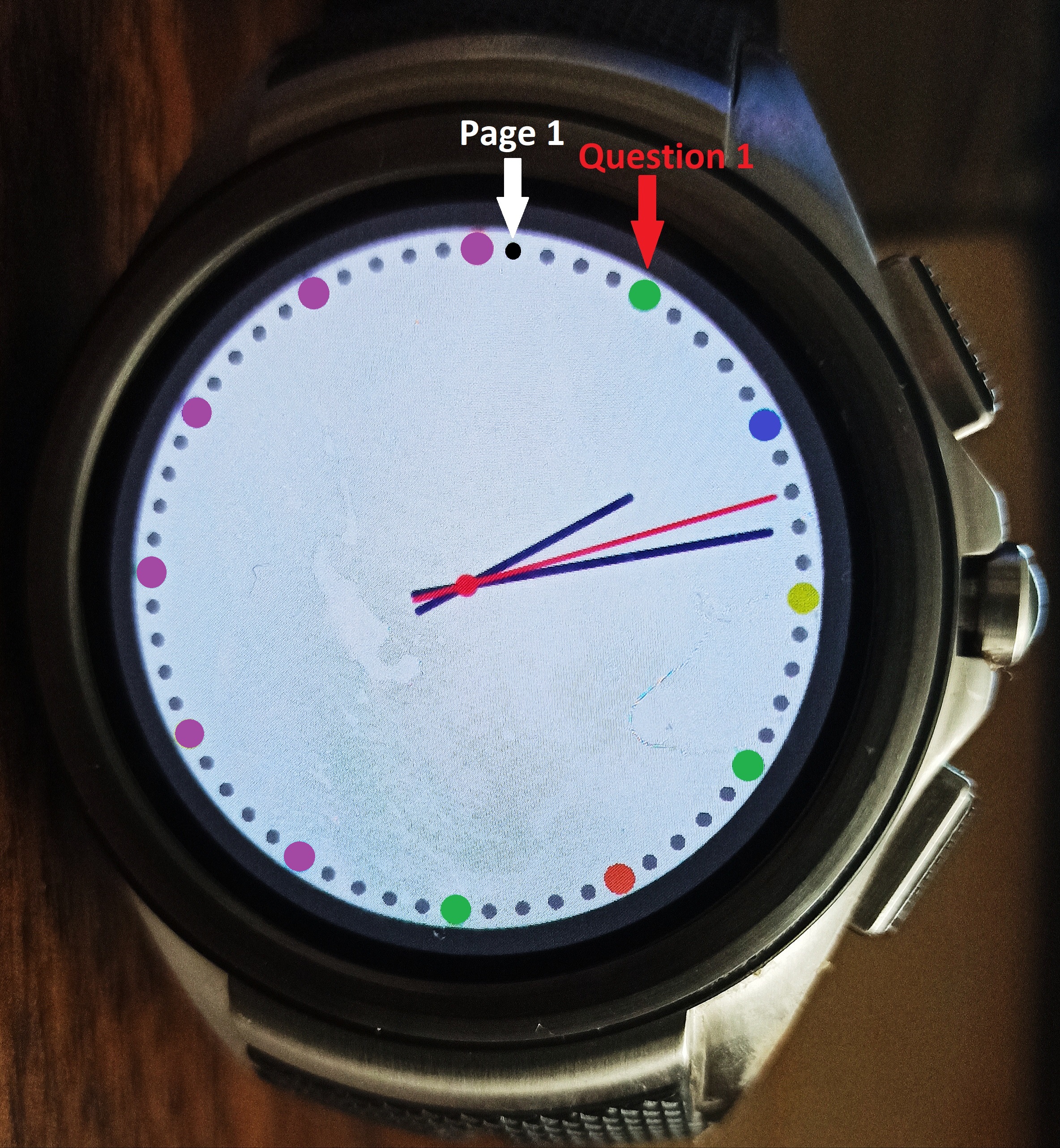}
    \caption{Stealthy visual interface for answer retrieval at beneficiary end of the attack.}
    \label{visual-2}
\end{wrapfigure}

Figure \ref{visual-2} shows the visual interface that we designed for answer retrieval by the beneficiary. The interface is obfuscated within an analog clock. The big dots represent the question numbers as well as the answers to each question. For example the red arrow points to the dot that represents Question 1. Question numbers increase clockwise with the maximum being Question 12 for {\it{a single page}}. The color of a dot represents the answer to the associated question. The answer options ``A'', ``B'', ``C'' and ``D'' are respectively represented by the colors red, green, blue and yellow, while a question that is yet to be answered is represented as a purple dot. For example, the answer to Question 1 is ``B'', while Questions 7 to 12 are yet to be answered.

The figure shows the answer status for {\it{a single page}} of 12 questions. For questions 13 and beyond, the small dots are used as a counter of page numbers. For example, in the figure the small dark grey dot (labelled Page 1 by the white arrow) implies that the screen shows the first page of 12 questions. For the next page of 12 questions, the current dark grey dot reverts to the default color (light grey) while the small dot next to it (clockwise) becomes dark grey. In general, the screen always has one small dark grey dot which determines the page (or the specific set of 12 questions). Since a typical exam would not have more than just a couple of sets of 12 questions, the small dark grey dot would in practice rarely go beyond the first 2-4 positions. Hence the beneficiary will easily see its location given a single glance on the screen, and easily work out the page number (and hence question numbers). To change the colors of the dots as new answers arrive we use the Android {\it{color class}} to assign the respective color constants to the various dot locations as needed. 

\subsubsection{Performance evaluation}
We recruited a group of 15 participants to evaluate the visual feedback system. The participants were first trained on how to retrieve the answers given the colored dots. Each of them was then later given the task to read a series of 15 answers. All participants found the scheme very intuitive and got all the answers correct. Our conclusion from this was that given a minimal amount of training, a beneficiary should in practice successfully use the system to retrieve the correct answers. 

\newtheorem{theorem}{Theorem}

\section{Theoretical Model of the Attack's Impact}
\label{model-1}

\subsection{Definitions and assumptions}
\label{def-1}
Let $N$ and $M$ be two positive integers. Assume an exam which {\it{has $N \geq 1$ multiple choice questions}}. Let each of these questions have {\it{$M>1$ answer options}} from which only one has to be selected as the correct answer. 
Assume that the mercenary chosen for the job has gestures recognized by the classification system with an accuracy of $\alpha$ on average, where 0$\leq$ $\alpha$ $\leq$ 1. We also alternately refer to $\alpha$ as the mercenary's skill (or simply, {\it{skill}}). Further assume that the mercenary's probability of answering a question correctly is $p$, where 0$\leq$ $p$ $\leq$ 1. We also alternately refer to $p$ as the mercenary's knowledge (or simply, {\it{knowledge}}). 

Let $X$ denote the random variable which defines the number of questions that the beneficiary will pass when our attack is used. The definition of $X$ assumes that the beneficiary will entirely rely on the answers provided by the attack, which means that {\it{he/she cannot override an answer provided by the system}}. This assumption enables our model to portray the impact of the attack independently of the beneficiary's potential intellectual contributions to solving the exam. For bench-marking purposes, we also define the random variable $Y$ which denotes the number of questions that the (same) student would have passed without cheating (i.e., if the attack had not been used). For ease of reference in our narrative we use the term {\it\underline{clean option}} to refer to the scenario when the student opts not to cheat. Let $\theta$ (0$\leq$ $\theta$ $\leq$ 1) represent the student's probability of passing an individual question under the {\it\underline{clean option}}. 

Finally, denote $r\leq N$ as the minimum number of questions that the student must get correct in order to pass the exam (i.e., $r$ is the pass-mark --- a score equal to or above $r$ means the student passes the exam). 



\begin{theorem}
Let $\beta=p\alpha+\frac{(1-p)(1-\alpha)}{(M-1)}$.    
The ratio $\mu$, of the probability that a beneficiary using our attack will pass the exam to the probability that the same student would have passed the exam under the clean option is given by
\begin{equation}
\label{eqn-1}
   \mu=\frac{\sum\limits_{X=r}^{N}{N \choose r}\beta^r (1-\beta)^{N-r}} {\sum\limits_{Y=r}^{N}{N \choose r}(\theta)^r(1-\theta)^{N-r}}
\end{equation}

\end{theorem}

For a proof of Theorem 1, see Appendix 1 (Page \pageref{app-1}). 
\subsection{Comparison to the weakest student ($\theta$=$\frac{1}{M}$)}

\begin{figure}
\centerline{
\includegraphics[width=0.39\textwidth]{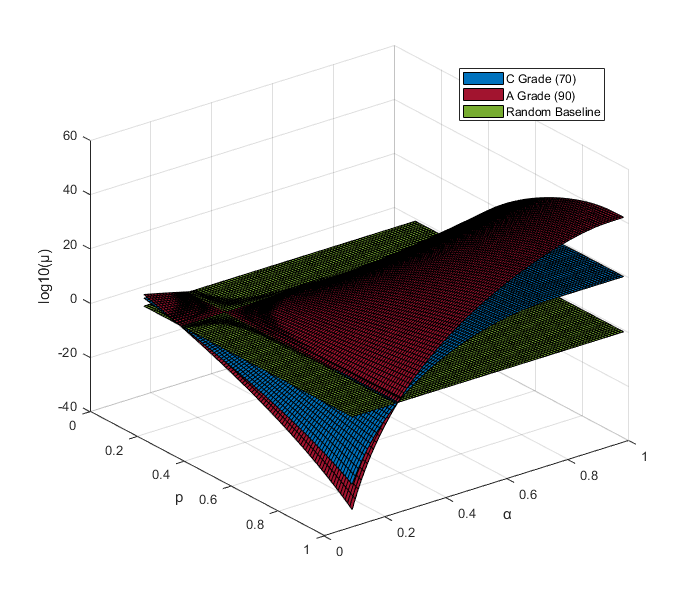}}
\caption{Illustration of the exam performance boost offered by our attack as a function of the mercenary's skill ($\alpha$) and knowledge ($p$). We use a logarithmic scale on the vertical axis to better visualize $\mu$ which spans many orders of magnitude.}
\label{fig:3dfig}
\vspace{-0.4 cm}
\end{figure}

A specific configuration of Theorem 1 that is of much significance for the characterization of our attack is that where $\theta$=$\frac{1}{M}$. Under this configuration, a student using the clean option is only able to {\it{randomly}} choose one of the $M$ answer options for each question. This configuration hence compares our attack to the worst possible student --- one who has no understanding at all of the material. 

Comparison with such a student enables us to determine the limits of the boost that our attack could theoretically ever provide to a student using it. Further, this kind of attack (or any form of cheating in general) is arguably most likely to be appealing to the weakest of students, further emphasizing why such students offer an interesting baseline. Figure \ref{fig:3dfig} visualizes Theorem 1 under this configuration ($\theta$=$\frac{1}{M}$). The figure shows how the boost $\mu$ (captured on a logarithmic scale) varies with the mercenary's skill ($\alpha$) and knowledge ($p$). We use $M$ = 4 to generate the plot since this is the most common configuration used in multiple choice exam questions. 

The dark red graph represents the scenario where the target pass grade is an ``A'' (i.e., the beneficiary seeks to pass a total of at least $r$= 90 questions out of $N$ = 100 questions on the exam). The blue graph on the other hand represents the scenario where the target pass grade is a ``C'' (i.e., the beneficiary seeks to pass a total of at least $r$= 70 questions out of $N$ = 100 questions on the exam). We use these two grade thresholds for our illustrations given their importance in the American education system. The former is the target for the high achieving student, while the latter is the one below which a student officially fails a course. The green plane cuts the other two graphs at the region where our attack performs similarly to the clean option (i.e., $\mu$=1 or $log10 (\mu)$=0)). We also refer to this plane as the random baseline because the particular instance of the clean option represented on the graph (i.e., $\theta$=$\frac{1}{M}$) is random as described earlier. The green plane is inserted to ease visualization of the impact of our attack relative to a student using the clean option.



{\bf{Main Observations}}: The overarching pattern depicted by Figure \ref{fig:3dfig} is that for both grade targets ``A'' and ``C'', our attack significantly increases the student's probability of passing the exam relative to the clean option for the biggest portion of the input space. For highly knowledgeable and skilled mercenaries (e.g., $\alpha$ = $p$ = 0.9), the attack is tens of orders of magnitude better than the clean option. Even for moderate values of mercenary skill and knowledge (e.g., $\alpha$ = $p$ = 0.4), the attack outperforms the clean option by at least couple of orders of magnitude. 
The overall conclusion drawn from the graph is that the attack is highly effective and that the mercenary does not have to have much skill or knowledge about course content for the attack to tremendously benefit the weak student. 

\begin{figure}[htbp]
\centerline{
\includegraphics[width=0.36\textwidth]{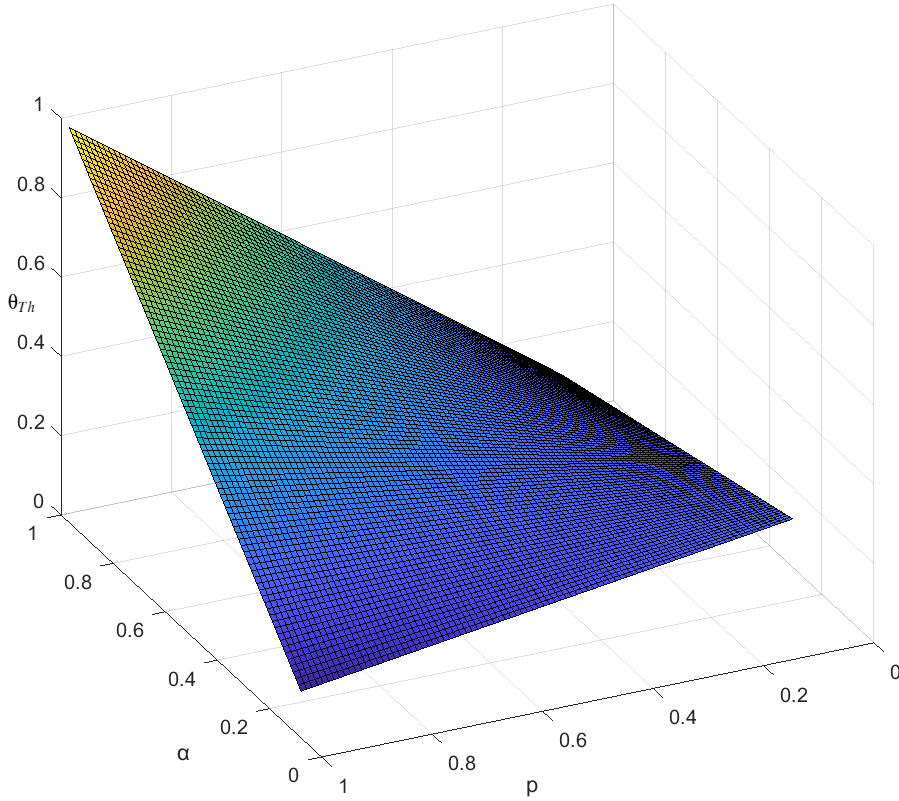}}
\caption{Illustration of the threshold that determines whether a student prefers our attack or the clean option.}
\label{fig:omega3D}
\vspace{-0.5 cm}
\end{figure} 

\subsection{Comparison to the general student}
In practice, students seeking to use our attack would have a wide range of capabilities under the clean option. This means that each student will have their own thresholds of attack specifications for which the attack (or clean option) would be preferred. The condition $\mu<1$ in Theorem 1 specifies the scenario when the student would have a higher probability of passing the exam under the clean option than when using our attack.  Exploiting the symmetry between the numerator and denominator of Equation 1, this condition simplifies down to  $\theta>\beta$. Hence the general student whose capabilities under the clean option map to some arbitrary value of $\theta$ will prefer the clean option if $\theta>$$p\alpha+\frac{(1-p)(1-\alpha)}{(M-1)}$ (i.e., prefer the attack if $\theta<$$p\alpha+\frac{(1-p)(1-\alpha)}{(M-1)}$). 

The first takeaway from this condition is that it is independent of $N$ or $r$. Hence a student seeking to leverage our attack to outperform the clean option needn't worry about the pass-mark or number of questions on the exam. All that matters  are the mercenary's skill and knowledge as well as the student's capabilities under the clean option and the number of answer options $M$, per question. Other key insights about the condition can be gleaned from the shape of the surface, $\theta_{Th}=p\alpha+\frac{(1-p)(1-\alpha)}{(M-1)}$ which defines the threshold $\theta=\theta_{Th}$, above which clean option is preferred to the attack. The surface is a hyperbolic paraboloid (see surface plotted in Figure \ref{fig:omega3D} for $M$=4) and thus has slopping behavior that varies  depending on the region of the surface under consideration. Taking the partial derivative with respect to $\alpha$ gives: $\frac{\partial\theta_{Th}}{\partial\alpha}$=$\frac{pM-1}{M-1}$. This derivative grows (linearly) with increasing $p$. The practical implication of this is that when $p$ is small (mercenary not so knowledgeable), small changes in $\alpha$ (e.g., through a little bit of gesture practice) yield modest changes in $\theta_{Th}$. But when when $p$ is large, small changes in $\alpha$ yield more significant changes in $\theta_{Th}$. 

\section{Conclusions}
\label{conc-1}
In this paper, we have showcased an attack which uses a smart watch as a tool to cheat examinations that use multiple choice questions. These types of questions form the core of many standardized tests and examinations in high schools and universities allover the world (see bottom of this page \cite{mcqwiki} for a list of prominent exams using these questions). 

The most obvious defence against this kind of attack is a randomization of questions in which each student has questions ordered differently. Several computer-based exams already do this. However, for paper-based exams which are graded by humans, such a change would make grading much more difficult. A more rigorous solution against these kinds of attacks seems to be the complete banning of phones and wearables in exams even when their usage appears benign to the plain eye.

\bibliographystyle{IEEEtran}
\bibliography{references}

\begin{thebibliography}{1}
\providecommand{\url}[1]{#1}
\csname url@samestyle\endcsname
\providecommand{\newblock}{\relax}
\providecommand{\bibinfo}[2]{#2}
\providecommand{\BIBentrySTDinterwordspacing}{\spaceskip=0pt\relax}
\providecommand{\BIBentryALTinterwordstretchfactor}{4}
\providecommand{\BIBentryALTinterwordspacing}{\spaceskip=\fontdimen2\font plus
\BIBentryALTinterwordstretchfactor\fontdimen3\font minus
  \fontdimen4\font\relax}
\providecommand{\BIBforeignlanguage}[2]{{%
\expandafter\ifx\csname l@#1\endcsname\relax
\typeout{** WARNING: IEEEtran.bst: No hyphenation pattern has been}%
\typeout{** loaded for the language `#1'. Using the pattern for}%
\typeout{** the default language instead.}%
\else
\language=\csname l@#1\endcsname
\fi
#2}}
\providecommand{\BIBdecl}{\relax}
\BIBdecl

\bibitem{migicovsky2014outsmarting}
A.~Migicovsky, Z.~Durumeric, J.~Ringenberg, and J.~A. Halderman, ``Outsmarting
  proctors with smartwatches: A case study on wearable computing security,'' in
  \emph{International Conference on Financial Cryptography and Data
  Security}.\hskip 1em plus 0.5em minus 0.4em\relax Springer, 2014, pp. 89--96.

\bibitem{wong2017assessing}
S.~Wong, L.~Yang, B.~Riecke, E.~Cramer, and C.~Neustaedter, ``Assessing the
  usability of smartwatches for academic cheating during exams,'' in
  \emph{Proceedings of the 19th international conference on human-computer
  interaction with mobile devices and services}, 2017, pp. 1--11.

\bibitem{mcqwiki}
``Multiple choice,'' Accessed on October 15, 2020,
  \url{https://en.wikipedia.org/wiki/Multiple\_choice}.

\end{thebibliography}

\appendices
\appendices
\section*{Appendix 1 -- Proof of Theorem 1}
\label{app-1}
\begin{proof}

Let $\beta$ (where 0 $\leq$ $\beta$ $\leq$ 1) represent the probability that the {\it{beneficiary}} selects the correct answer to a question. The random variable 
$X$ is modeled by a binomial distribution with parameters $\beta$ and $N$ (i.e., $X$$\sim$ $Bin$ ($\beta$, $N$)). Using similar reasoning for the student using the clean option, it follows that $Y$$\sim$ $Bin$ ($\theta$, $N$). 
 
The probability that the beneficiary passes the exam, is thus given by P(X$\geq$ r) = $\sum\limits_{X=r}^{N}{N \choose r}\beta^r (1-\beta)^{N-r}$, while the probability that the student using the clean option will pass the exam is given by $P(Y\geq r) = \sum\limits_{Y=r}^{N}{N \choose r}(\theta)^r(1-\theta)^{N-r}$. Dividing these two expressions gives the formulation for $\mu$ shown in Equation \ref{eqn-1}. To complete the proof, we next find $\beta$ in terms of our core problem variables. Assuming that the beneficiary entirely relies on the mercenary and the associated attack framework for all answers, there are two ways in which the beneficiary can pass a given question. These are:

{\bf{Scenario \# 1}}: The mercenary selects a correct answer (with probability $p$) AND the gesture recognition system correctly classifies the mercenary's wrist movements mapping to this answer (with probability $\alpha$). Using the product rule, the probability of Scenario \# 1 is thus $p$$\alpha$.

{\bf{OR}},

{\bf{Scenario \# 2}}: The mercenary selects a wrong answer (with probability $1-p$) AND the gesture recognition system {\it{incorrectly}} classifies the mercenary's wrist movements in such a way that the correct answer ends up being selected (with probability $\frac{(1-\alpha)}{(M-1)}$). The $\frac{(1-\alpha)}{(M-1)}$ term is obtained as follows. If out of $M$ options the mercenary selects one (wrong) answer, this leaves behind $M$-1 options. Since we know that the classification system makes an error, this means it does not select the (wrong) option selected by the mercenary. Rather, it selects one of the $M$-1 options not selected by the mercenary. But we know that one of these $M$-1 options is the correct answer. Hence the probability that the classifier selects this single correct answer out of the $M$-1 options is $\frac{(1-\alpha)}{(M-1)}$ (i.e., the idea is to share the mis-classification probability (1-$\alpha$) equally between the $M$-1 remaining options).  Using the product rule, the probability of Scenario \# 2 is thus $\frac{(1-p)(1-\alpha)}{(M-1)}$. By adding the results from Scenarios \# 1 and \# 2, the probability $\beta$ that the beneficiary selects the correct answer to a question is given by 

\begin{equation*}
\beta=p\alpha+\frac{(1-p)(1-\alpha)}{(M-1)}    
\end{equation*} 

which completes our proof.

\end{proof}

\ifCLASSOPTIONcaptionsoff
  \newpage
\fi

\end{document}